\documentclass{iopart}
\usepackage{amsmath,amssymb,graphicx}
\usepackage{epsfig,color}
\newcommand{\bmsigma}{\boldsymbol \sigma}

\newcommand\cmr{\bmsigma_{\!\varrho}}

\def\S{{\sf S}}
\def\sm{{\scriptstyle \!S}}

\def\X{\boldsymbol{X}}
\def\Tr{\hbox{Tr}} 
\newtheorem{tth}{Theorem}
\begin{document}
\title{Geometry of perturbed Gaussian states and quantum estimation}
\author{
Marco G. Genoni\footnote[1]{\tt m.genoni@imperial.ac.uk}$^{1}$
Paolo Giorda\footnote[2]{\tt giorda@isi.it}$^{2}$
Matteo G A Paris\footnote[5]{\tt matteo.paris@fisica.unimi.it}$^{3}$
}
\address{
$^1$QOLS, Blackett Laboratory, Imperial College London, London SW7 2BW, UK \\
$^2$Institute for Scientific Interchange Foundation, I-10133 Torino, Italy\\
$^3$Dipartimento di Fisica, Universit\'a degli Studi di Milano, I-20133
Milano, Italy}
\begin{abstract}
We address the nonGaussianity (nG) of states obtained by weakly
perturbing a Gaussian state and investigate the relationships with
quantum estimation. For classical perturbations, i.e. perturbations to
eigenvalues, we found that nG of the perturbed state may be written as
the quantum Fisher information (QFI) distance minus a term depending on
the infinitesimal energy change, i.e. it provides a lower bound to
statistical distinguishability.  Upon moving on isoenergetic surfaces in
a neighbourhood of a Gaussian state, nG thus coincides with a proper
distance in the Hilbert space and exactly quantifies the statistical
distinguishability of the perturbations. On the other hand, for
perturbations leaving the covariance matrix unperturbed we show that nG
provides an upper bound to the QFI.  Our results show that the geometry
of nonGaussian states in the neighbourhood of a Gaussian state is
definitely not trivial and cannot be subsumed by a differential
structure.  Nevertheless, the analysis of perturbations to a Gaussian
state reveals that nG may be a resource for quantum estimation. The nG
of specific families of perturbed Gaussian states is analyzed in some
details with the aim of finding the maximally non Gaussian state
obtainable from a given Gaussian one.  \end{abstract}
\section{Introduction}
NonGaussianity (nG) is a resource for the implementation of continuous
variable quantum information in bosonic systems \cite{ngl10}. 
Several schemes to generate nonGaussian states from Gaussian ones 
have been proposed,
either based on nonlinear interactions or on conditional measurements
\cite{DeGaussTh1,DeGaussTh2,Lvo01,Wen04,Zav04,Our06a,
Nee07,Our07a,Zav07a,Par07,Zav09,Our09,Our07,DistTaka,ngOPO1,ngOPO2,
kerr:PRL:01,kerr:PRA:06,kerr:NJP:08,mas06,ps10,Therm,MTWBRD,PhAdd}.
In many cases the effective nonlinearity is small, and so it
is the resulting nG. It is thus of interest to investigate 
the nG of states in the neighbourhood of a Gaussian state,
i.e. the nG of slightly perturbed Gaussian states. Besides the
fundamental interest \cite{All10} this also provides a way to assess different
deGaussification mechanisms, as well as nG itself as a resource 
for quantum estimation. Indeed, in an estimation problem where the 
variation of a parameter affects the Gaussian character of the involved
states one may expect the amount of nG to play a role in 
determining the estimation
precision. 
\par
Quantum estimation deals with situations where
one tries to infer the value 
of a parameter $\lambda$ by measuring a different quantity $X$, which 
is somehow related to $\lambda$. This often happens in quantum mechanics 
and quantum information where many quantities of interest, e.g. entanglement
\cite{EE,EEE}, do not correspond to a proper observable and should be
estimated from the measurement of one or more observable
quantities \cite{LQE}. Given a set $\{\varrho_\lambda\}$ of quantum 
states parametrized by the value of the quantity of interest, an estimator 
$\hat\lambda$ for $\lambda$ is a real function of the outcomes of the 
measurements performed on $\varrho_\lambda$. The quantum Cramer-Rao 
theorem \cite{LQE1,LQE2,LQE3,LQE4} establishes a lower bound for the variance
${\mathrm{Var}}(\lambda)$ of any unbiased estimator, i.e. for the 
estimation precision,  
$
{\mathrm{Var}}(\lambda) \geq (M H(\lambda))^{-1}
$
in terms of the number of measurements $M$ and the so-called quantum Fisher 
information (QFI), which captures the statistical distinguishability of the
states within the set. Indeed, the QFI distance itself is proportional to 
the Bures distance $d^2_B (\varrho_1,\varrho_2)=2[1-F(\varrho_1,\varrho_2)]$, 
$F(\varrho_1,\varrho_2)=\hbox{Tr}[\sqrt{\sqrt{\varrho_1}\varrho_2\sqrt{\varrho_2}}]$ 
being the fidelity, between 
states corresponding to infinitesimally close values of the parameter,
i.e., in terms of metrics,
$H(\lambda)  = 4\,g_\lambda
= 2 \sum_{nm} (\varrho_n+ \varrho_m)^{-1}\left|\langle \psi_m| \partial_\lambda
\varrho_\lambda | \psi_n\rangle\right|^2 $
where
$d^2_B (\varrho_{\lambda+d\lambda},\varrho_\lambda)=g_\lambda\,
d\lambda^2$, and we have used the eigenbasis $\varrho_\lambda = \sum_n \varrho_n 
|\psi_n\rangle\langle\psi_n |$. 
\section{Gaussian states and a measure of non Gaussianity}
Let us consider a single-mode bosonic system described by the mode 
operator $a$ with commutation relations $[a,a^\dag]=1$. A quantum
state $\varrho$ is fully described by its characteristic function 
$\chi[\varrho](\lambda) = \Tr[\varrho\:D(\lambda)]$ where
$D(\lambda) =\exp\{\lambda a^{\dag} - \lambda^* a \}$ is the
displacement operator.  The canonical operators are given by
$q = (a + a^{\dag})/\sqrt{2}$ and
$p = (a - a^{\dag})/\sqrt{2}i$
with commutation relations given by $[q,p]=i$.
Upon introducing the vector $\boldsymbol{R}=(q,p)^T$,
the covariance matrix $\cmr$ and the vector
of mean values $\X_{\!\varrho}$ of a quantum state
$\varrho$ are defined as
$\sigma_{kj} =\frac{1}{2}\langle R_k R_j + R_j R_k\rangle_\varrho - 
\langle R_j\rangle_\varrho\langle R_k\rangle_\varrho$ and
${X}_j = \langle R_j \rangle_\varrho$,
where
$\langle O \rangle_\varrho = \Tr[\varrho\,O]$ is the expectation value
of the operator $O$ on the state $\varrho$.
A quantum state
is said to be Gaussian if
its characteristic function has a Gaussian form. 
Once the CM and
the vectors of mean values are given, a Gaussian state is fully
determined.  
\par
The amount of nG $\delta[\varrho]$ of a quantum state 
$\varrho$ may be quantified by the quantum relative entropy \cite{nge} 
$\S(\varrho \rVert \tau_\varrho)=\Tr[\varrho (\log \varrho - \log\tau)]$
between $\varrho$ and its reference Gaussian state $\tau_\varrho$,
which is a Gaussian state with the same covariance 
matrix $\bmsigma$ as $\varrho$.
As for its classical counterpart, the Kullback-Leibler divergence, it can
be demonstrated that $0\leq \S(\varrho \rVert \tau) < \infty$ when it
is definite, \emph{i.e.} when
$\mathrm{supp} \: \varrho \subseteq \: \mathrm{supp} \: \tau$.
In particular $\S(\varrho \rVert \tau_\varrho)=0$ iff
$\varrho\equiv\tau_\varrho$  
\cite{SchumacherRelEnt,VedralRelEnt}.
Since $\tau_\varrho$ is Gaussian
$\Tr[(\tau_\varrho -\varrho)\: \log\tau_\varrho]=0$, {\em i.e.}
$\S(\varrho\rVert\tau_\varrho) = \S(\tau_\varrho) - \S(\varrho)$ 
and we may write
$\delta[\varrho] = \S(\tau_\varrho) - \S(\varrho)$ where
$\S(\varrho) = - \Tr[ \varrho \: \log \varrho ]$ is the von Neumann entropy of
$\varrho$. Finally, since the von Neumann entropy of
a single-mode Gaussian state may be
written as $h(\sqrt{\det \cmr})$ where $h(x) = (x+\frac12)
\log(x+\frac12))-(x-\frac12) \log(x-\frac12)$ we have
\begin{align}
\delta[\varrho]=h(\sqrt{\det\cmr})-\S(\varrho)\,.
\label{ng}
\end{align}
\par
A generic single-mode Gaussian state may be written as
$\tau=U_\sm\, \nu\, U^\dag_\sm$ where $U_\sm$ is a symplectic operation i.e. a 
unitary $U_\sm=\exp{-iH}$
resulting from a Hamiltonian $H$ at most quadratic in the field
operators, and $\nu$ is a chaotic (maximum entropy) state with 
$n_\nu=\hbox{Tr}[\nu\, a^\dag a]$ average thermal
quanta, i.e.  $\nu=\sum_k p_k |k\rangle\langle k|$ , $p_k =
n_\nu^k/(1+n_\nu)^{1+k}$ in the Fock number basis. 
\section{Classical perturbations to a Gaussian state}
An infinitesimal 
perturbation of the eigenvalues $p_k$ of a Gaussian state $\tau$, i.e.
$
p_k \rightarrow p_k+ dp_k 
$
results in a perturbed state $\varrho=\sum_k (p_k +
dp_k) U_\sm|k\rangle\langle k|U^\dag_\sm$ which, in general, 
is no longer Gaussian. 
Since the nG of a state is invariant 
under symplectic operations we have $\delta[\varrho]=\delta[\eta]$
where $\eta = U^{\dag}_\sm\! \varrho\, U_\sm = \sum_k (p_k +
dp_k) |k\rangle\langle k|$ is diagonal in the Fock basis. 
The Gaussian reference $\tau_\eta$ of $\eta$ is a thermal state with
$n_\eta=n_\nu+dn=\sum_k k\,(p_k + dp_k)$ average quanta and the nG
may be evaluated upon expanding both terms in
$\delta[\eta](=\delta[\varrho])$
up to the second order, 
\begin{align}
\delta[\varrho] = 
\sum_k \frac{dp_k^2}{2 p_k} - \frac{dn^2}{2 n_\nu(1+n_\nu)}\,. 
\label{eq:nonGBures}
\end{align}
NonGaussianity of perturbed states is thus given by the sum of two
contributions. The first term is the Fisher information of the probability
distribution $\{p_k\}$, which coincides with the classical part of the Bures 
distance in the Hilbert space. The second term is a negative
contribution expressed in terms of the infinitesimal change of the
average number of quanta. When traveling on surfaces at constant 
energy the amount of nG coincides with a proper distance
in the Hilbert space and, in this case, it has a geometrical 
interpretation as the infinitesimal Bures distance. At the same time,
since Bures distance is proportional to the QFI one, it expresses the 
statistical distinguishability of states, and we conclude that moving 
out from a Gaussian state towards its
nonGaussian neighbours is a resource for estimation purposes.
Similar conclusions can be made when comparing families of perturbations 
$\{ dp \}$ corresponding to the same infinitesimal change of energy 
$dn^2$: in this case the different amounts of  nG induced 
by the perturbations are quantified by the Bures distance minus a constant 
term depending on $dn^2$ and the initial thermal energy $n_\nu$, i.e.
the intial purity $\mu\equiv\hbox{Tr}[\varrho^2] = \hbox{Tr}[\tau^2]=
(2 n_\nu +1)^{-1}$
We summarize the above statements in the following
\begin{tth}
If $\tau_\lambda$ is a Gaussian state and an infinitesimal variation of
the value of $\lambda$ drives it into a state $\varrho_{\lambda+d\lambda}$ 
with the same eigenvectors, then the QFI distance 
is equal 
to the nG $\delta[\varrho_{\lambda+d\lambda}]$ plus a term
depending both on the infinitesimal variation of energy $dn^2$ and
on the initial purity
$$ H(\lambda) d\lambda^2 = \delta[\varrho_{\lambda+d\lambda}] + 
\frac{2\, \mu^2 dn^2}{1-\mu^2}
\:.
$$
In particular, for perturbations that leave the energy unperturbed 
the nG of the perturbed state coincides with 
the QFI distance, whereas, in general, it provides a 
lower bound. $\square$
\end{tth}
\subsection{Examples of finite perturbations}
In order to explore specific directions in the neighbourhood of 
a Gaussian state $\tau$ let us write the perturbation to the 
eigenvalues as $dp_k= \epsilon \mu_k$ 
where $\{\mu_k\}$ is a given distribution. In this case the
nG of the perturbed state is given by
\begin{align}
\delta[\varrho] = \epsilon^2\, \left(\sum_k 
\frac{(p_k-\mu_k)^2}{2 p_k} -
\frac{\Delta n_\mu^2}{2 n_t(1+n_t)}\right) 
+ O(\epsilon^3)\,,\label{eq:nonGExp}
\end{align}
where
$
\Delta n_\mu = \sum_k (p_k - \mu_k) \langle k|  a^\dag a |k\rangle 
$
Let us now consider the families of states generated by the 
convex combination $\varrho = (1-\epsilon)\tau
+ \epsilon \varrho_\mu$ of the Gaussian states 
$\tau$ with a the target state 
$\varrho_\mu=\sum_k \mu_k U_\sm|k\rangle\langle k|U^\dag_\sm$, which
itself is  obtained by changing the eigenvalues $p_k$ of the initial 
Gaussian state to $\mu_k$. Again we exploit invariance of
$\delta[\varrho]$ under symplectic operations and focus attention to 
the diagonal state $\eta = U_\sm^{\dag}\varrho U_\sm=(1-\epsilon)\eta
+\epsilon\, \eta_\mu$ which has the same nG of $\varrho$.
This is the generalization to a finite perturbation of the analysis reported 
in the previous Section, and it is intended as a mean to find the maximally nonGaussian state 
obtainable starting from a given Gaussian. NonGaussianity of this kind of 
states can be written as
$\delta[\eta] = S(\tau_\eta) - S(\eta) 
= h[n_\eta + 1/2] - H[q^{(\epsilon)}] 
$, where  $H[q^{(\epsilon)}]=-\sum_k q^{(\epsilon)}_k \log q^{(\epsilon)}_k$
denotes the Shannon entropy of the distribution $q_k^{(\epsilon)} =
(1-\epsilon) p_k + \epsilon \mu_k$ and  
$n_\eta =\Tr[\eta \, a^{\dag} a]= (1-\epsilon) n_t + \epsilon\, n_\mu$ 
is the average number of quanta of $\eta$.
Notice that for a thermal state with $n_t$ quanta
we have ${H}[p] = h[n_t + 1/2]$.
By using the 
concavity of the Shannon entropy we obtain an upper bound
for the nG
\begin{align}
\delta[\eta] \leq h[n_\eta + 1/2] - h[n_t + 1/2] +
\epsilon \Big( h[n_t + 1/2] - H[\mu] \Big)
\end{align}
In particular, if the two distributions have the same number of quanta, 
$n_\mu=n_t$, and thus $n_t=n_\eta$, the bound on $\delta[\eta]$ only 
depends on the difference between the entropy of the initial and target 
distributions.  
\par
Let us now consider perturbations towards some relevant distributions
i.e. Poissonian $\mu_{k}^{(p)} =
\frac{n_\mu^k}{k!}e^{-n_\mu}$, thermal 
$\mu_k^{(t)}$, and Fock 
$\mu_k^{(d)} =\delta_{k,n_\mu}$ 
and evaluate the nG of states obtained as convex combination 
of a thermal state with $n_t$ quanta and a diagonal quantum state with a 
Poissonian, thermal or Fock distributions and $n_\mu$ quanta.
\begin{figure}[ht]
\begin{tabular}{cc} 
\includegraphics[width=0.35\columnwidth]{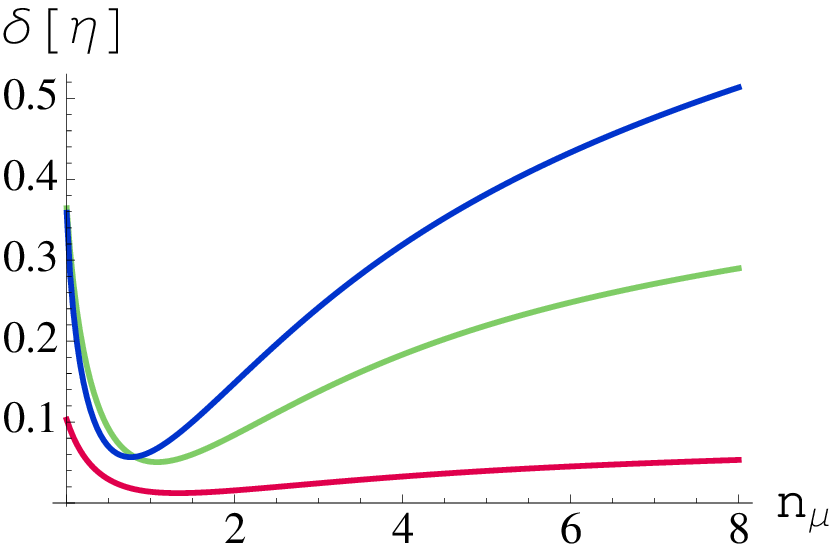} 
& \includegraphics[width=0.35\columnwidth]{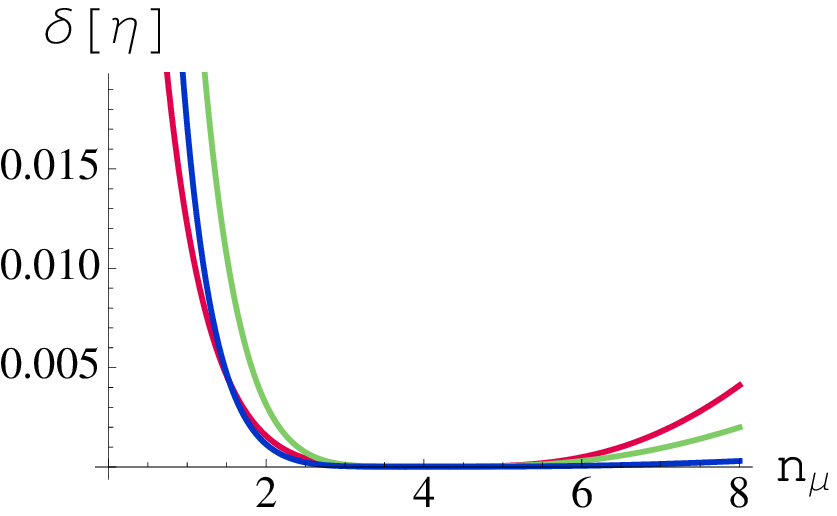} 
\\ \includegraphics[width=0.35\columnwidth]{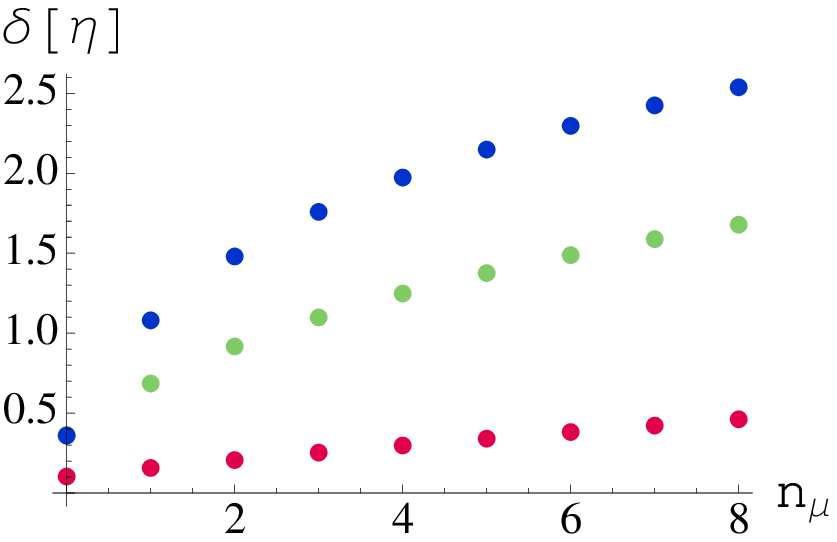} 
& \includegraphics[width=0.35\columnwidth]{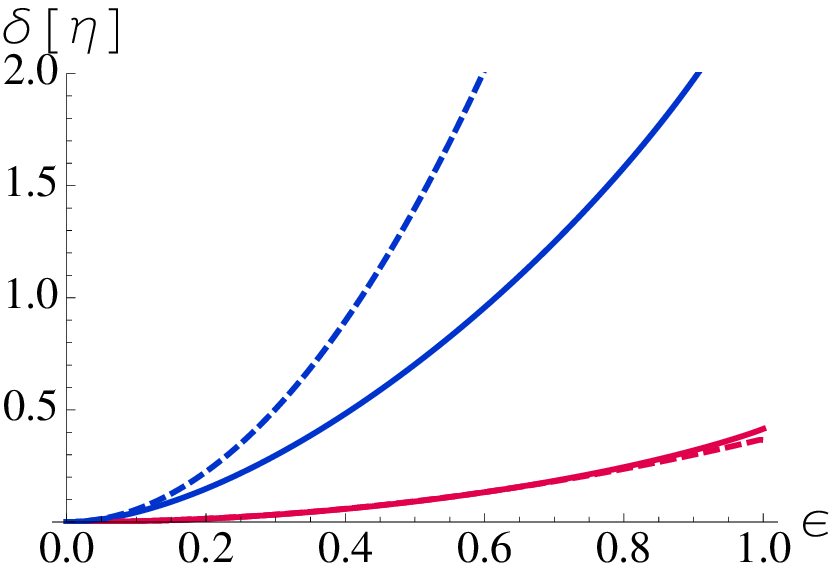} 
\end{tabular}
\caption{NonGaussianity $\delta[\eta]$ as a function of the
energy of the target state $n_\mu$ , with $n_t=4$ and
for different values of $\epsilon=0.3, 0.7, 0.9$ (red, green and 
blue, respectively) for Poissonian target (upper left panel), 
thermal (upper right) and Fock (lower left). 
The lower right panel shows the nG $\delta[\eta]$ as a function of 
$\epsilon$, with $n_\mu=n_t=4$ and
for convex combination with a Poissonian (red) and a Fock (blue) 
state. Dashed lines correspond to the expansion of nG at 
the second order in $\epsilon$, as obtained in Eq. (\ref{eq:nonGExp}) for
the corresponding perturbations.
\label{f:1n}}
\end{figure}
\par
In Fig. \ref{f:1n} we plot the nG of the convex combination $\varrho$ as
a function of $n_\mu$ for different values of $\epsilon$: If we consider
the convex combination with a Fock state, the nG simply increases
monotonically with the energy of the added state.  For combinations with
Poissonian and thermal distributions we have a maximum for $n_\mu
\rightarrow 0$, then a local minimum (which for the thermal distribution
corresponds trivially to $\delta[\varrho]=0$) for $n_\mu=n_t$, and then
the nonG increases again for higher values of $n_\lambda$. This implies
that in order to increase nG the best thing to do is to perturb the
initial Gaussian either with a highly excited state $n_\mu \gg 1$ or
with the vacuum state $|0\rangle\langle 0|$.
If we choose  $n_\varrho=n_\mu = n_t$ 
we know from Eq. (\ref{eq:nonGBures}) that nG for small 
perturbations is equivalent to the Bures infinitesimal distance between 
the two probability distributions ${\mu_k}$ and ${p_k}$. In  Fig. 
\ref{f:1n}d we show the nG as a function of $\epsilon$: 
As it is apparent from the plot the nG obtained 
by adding a Fock state is always much larger that 
the one for a Poissonian profile. 
We observe that in the latter case the expansion at the second
order obtained in Eq. (\ref{eq:nonGExp}) is still accurate for values
of $\epsilon$ approaching $1$, while it fails to be accurate for 
$\epsilon \gtrsim 0.2$ for a Fock state, becoming an upper bound
on the exact amount of nG. We have also investigated
what happens by considering a target distribution randomly chosen on
a finite subspace of the infinite Hilbert space: Again we have obtained
that, at fixed energy of the target state, perturbing with a Fock state
yields the biggest increase of nG. This may be easily generalized to a 
system of $d$ bosonic modes, where the most general Gaussian state is 
described by $d(2d +3)$
independent parameters.
\section{Perturbations at fixed covariance matrix}
Gaussian states are known to be
extremal state at fixed covariance matrix 
for several relevant quantities, e.g. channel capacities and entanglement 
measures \cite{Wolf}. Therefore, one may wonder
whether perturbing a Gaussian state at fixed covariance matrix may be
quantified in convenient way for the purposes of quantum estimation.
This indeed the case: the nG provides an upper bound 
to the QFI distance at fixed covariance matrix and thus
have an operational interpretation in terms of statistical distinguishability. 
This is more precisely expressed by the following theorem \cite{ngl10}
\begin{tth}
If $\tau_\lambda$ is a Gaussian state and an infinitesimal variation of
the value of $\lambda$ drives it into a state $\varrho_{\lambda+d\lambda}$ 
with the same covariance matrix, then the nG 
$\delta[\varrho_{\lambda+d\lambda}]$ provides an upper bound to the
QFI distance
$$ H(\lambda) d\lambda^2 \leq \delta[\varrho_{\lambda+d\lambda}] \:.
$$
\end{tth}
{\bf Proof}: If $\varrho_{\lambda+d\lambda}$ and $\tau_\lambda$ have
the same CM then the nG of $\varrho_{\lambda+d\lambda}$, 
$\delta[\varrho_{\lambda+d\lambda}] =
\S(\varrho_{\lambda+d\lambda}||\tau_\lambda)=\widetilde
H(\lambda)d\lambda^2$, where the so-called
Kubo-Mori-Bogolubov information $\widetilde H(\lambda)$ \cite{WHH07,Ama00}
provides an upper bound for the quantum Fisher information 
$H(\lambda) \leq \widetilde H (\lambda)$ \cite{Pet96}, thus proving the theorem.
$\square$
\\ $ $ \par
The above theorem says that a larger nG of the perturbed state 
may correspond to a greater distinguishability from the original 
one, thus allowing  a more precise estimation. Of course, this is
not ensured by the theorem, which only provides an upper bound
to the QFI. One may wonder that when
$\varrho_{\lambda+d\lambda}$ is itself a Gaussian state
the theorem requires $H(\lambda)=0$, i.e. no reliable estimation
is possible. Indeed, this should be the case, since Gaussian states
are uniquely determined by the first two moments and thus the 
requirement that the perturbed $\tau_{\lambda+d\lambda}$ and the 
original state $\tau_{\lambda}$ are both Gaussian and have the same 
covariance matrix implies that they are actually the same quantum state.
\section{Conclusions}
In conclusion, we have addressed the nG of states obtained by weakly
perturbing a Gaussian states and have investigated the relationships
with quantum estimation.  We found that nG provides a lower bound to the
QFI distance for classical perturbations, i.e. perturbations to
eigenvalues leaving the eigenvectors unperturbed, and an upper bound for
perturbations leaving the covariance matrix unperturbed.  For situations
where the CM is changed by the perturbation we have no general results.
On the other hand, it has been already shown that non-Gaussian states
improve quantum estimation of both unitary perturbations as the
displacement and the squeezing parameters \cite{EKerr} and nonunitary
ones as the loss parameter of a dissipative channel \cite{GG09}.
Overall, our results show that the geometry of nonGaussian states in the
neighbourhood of a Gaussian state is definitely not trivial and cannot
be subsumed by a differential structure. Despite this fact, the analysis
of perturbations to a Gaussian state may help in revealing when, and to
which extent, nG is a resource for quantum estimation. We have also
analyzed the nG of specific families of perturbed Gaussian states with
the aim of finding the maximally non Gaussian state obtainable from a
given Gaussian one.
\ack
MGG acknowledge the UK EPSRC for financial support.
MGAP thanks R. F. Antoni for being a continuing inspiration.

\end{document}